\documentclass[a4paper,11pt]{article}
\usepackage{jheppub} 
\usepackage[capitalize]{cleveref}
\usepackage{amsmath}
\usepackage{bbold}
\usepackage{appendix}
\usepackage[dvipsnames]{xcolor}
\usepackage[normalem]{ulem}
\usepackage{tikz}
\usetikzlibrary{decorations.pathmorphing}\usetikzlibrary{positioning,calc}
\usetikzlibrary{math}
\graphicspath{{plots/}}

\newcommand{\B}{b}
\newcommand{\eV}{\textrm{eV}}

\newcommand{\GeV}{\textrm{GeV}}
\newcommand{\TeV}{\textrm{TeV}}
\newcommand{\DM}{\textsc{dm}}
\newcommand{\SM}{\textsc{sm}}

\newcommand{\OGW}{\Omega_\textsc{gw}}
\newcommand{\TRH}{T_\textsc{rh}}
\newcommand{\gs}{g_{\star s}}

\newcommand{\Hinf}{H_\text{inf}}
\newcommand{\Tinf}{T_\text{inf}}
\newcommand{\Tnuc}{T_\text{nuc}}
\newcommand{\Trh}{T_\textsc{rh}}
\newcommand{\Tfo}{T_\text{f.o.}}

\newcommand{\fpk}{f_\text{pk}}

\newcommand{\LQCD}{\Lambda_\textsc{qcd}}
\newcommand{\LCDM}{\Lambda\text{CDM}}

\preprint{ZU-TH 86/25}

\title{%
Impact of Supercooling on Direct Searches for Dark Matter and Gravitational Wave Backgrounds%
}

\author[1,2]{Davide Racco,}
\author[3,4]{Alfredo Stanzione}

\affiliation[1]{Institut f\"ur Theoretische Physik, ETH Z\"urich, Wolfgang-Pauli-Str. 27, 8093 Z\"urich, Switzerland}
\affiliation[2]{Physik-Institut, Universit\"at Z\"urich, Winterthurerstrasse 190, 8057 Z\"urich, Switzerland}
\affiliation[3]{SISSA International School for Advanced Studies, Via Bonomea 265, 34136, Trieste, Italy}
\affiliation[4]{Università degli Studi di Roma la Sapienza, Piazzale Aldo Moro 5, 00185, Roma, Italy}
\emailAdd{davide.racco@physik.uzh.ch}
\emailAdd{alfredo.stanzione@sissa.it}

\abstract{
An interesting feature of a cosmological phase transition can be a stage of exponential expansion (supercooling).
The modified expansion history and the entropy injection at reheating, can affect the final energy fraction of dark matter.
In this paper, we revisit the calculation of the freeze-out and freeze-in dynamics, showing additional effects on top of the standard dilution factor if the dark matter production is completed during the supercooling stage.
We show for the first time how these effects can be particularly interesting for direct detection, as the parameter space for WIMP-like candidates shifts from excluded to allowed regions, and freeze-in candidates get closer to experimental reach.
A phenomenological motivation to consider supercooling is the associated gravitational wave background. 
The implications of a finite-duration reheating stage, when the equation of state is close to matter-domination, are a peculiar low-frequency spectrum, and its shift to lower frequencies. 
These effects are a complementary test of the dynamics that we study for dark matter production, and remarkably can link direct detection of dark matter and gravitational wave astronomy.
}

\begin{document}
\maketitle

\section{Introduction}

The existence of Dark Matter (DM) is among the leading pieces of evidence of physics beyond the Standard Model (BSM), and discovering its nature would have deep implications for particle physics.
As a guidance among the many possible particle candidates for DM, we might simultaneously address another puzzle of the SM (e.g.~Higgs mass hierarchy, strong CP problem, neutrino masses) or identify candidates whose abundance can be naturally achieved within the cosmological history.

The extrapolation of the SM to primordial epochs predicts a radiation-dominated phase (RD) from the end of primordial inflation until the equality epoch at $T\sim \eV$. 
The many open puzzles of the SM guarantee a rich BSM phenomenology, which would likely imply novel dynamics in the cosmological history with respect to the standard $\LCDM$. 
Two effects which could easily affect the predicted DM abundance are an alternative expansion history with an equation-of-state (EOS) parameter $w\neq \tfrac 13$, or an entropy injection (due to decay of non-relativistic species into the relativistic bath) \cite{Allahverdi:2020bys}. Remarkably, both effects are already present within $\LCDM$, respectively around the QCD transition, and throughout the thermal history whenever the bath temperature cools below the mass of some SM particle.

Both these cosmological phenomena are guaranteed to occur if there is a stage of supercooling (SC), which is plausible in the phase transitions (PT) associated to many BSM extensions.
The SM already features two PTs, which happen to be cross-overs with a moderate \cite{Laine:2015kra,Borsanyi:2016ksw} (but not negligible, see e.g.~\cite{Franciolini:2023wjm}) impact on the EOS parameter.

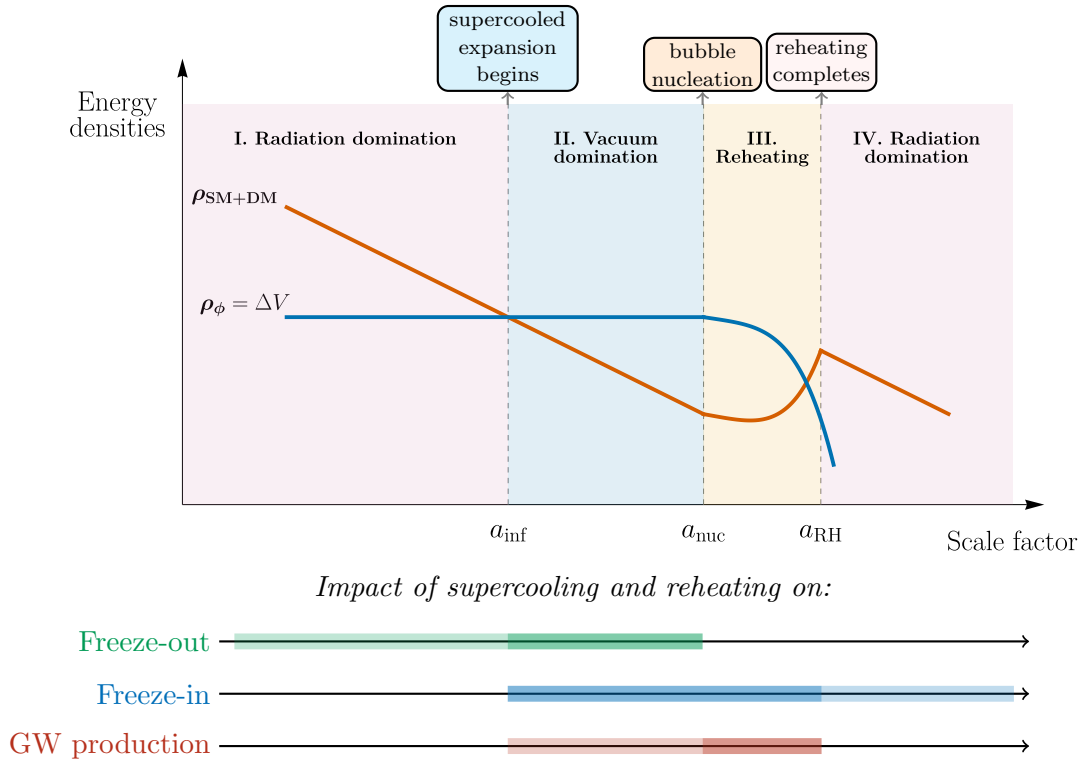
\begin{figure}\centering
\vspace*{-2em}
\begin{tikzpicture}
\node [above right,inner sep=0] (image) at (0,0) {\includegraphics[width=0.88\textwidth]{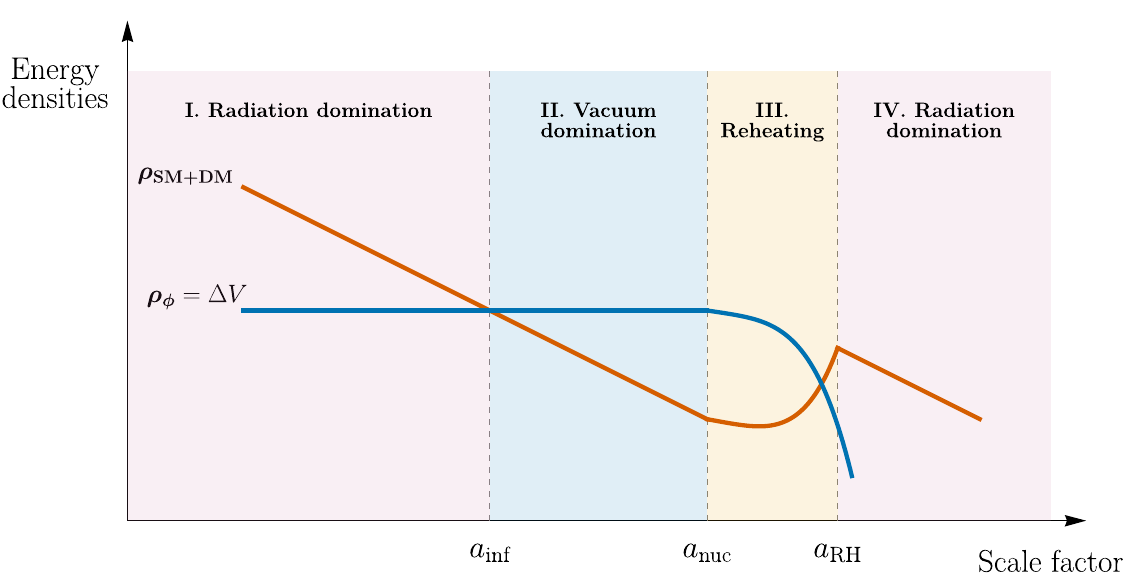}};
\begin{scope}[
x={($0.05*(image.south east)$)},
y={($0.1*(image.north west)$)}]
\def\ytitle{-0.5}
\def\yFO{-1.5}
\def\yFI{-2.5}
\def\yGW{-3.5}
\def\tinf{8.7}
\def\tnuc{12.55}
\def\trh{14.9}
\def\opa{.5}
\def\opb{.25}
\def\ylabel{9.0}
\begin{scriptsize}
\node[above, align=center, text width=1.7cm, minimum height=2em,inner sep=2pt,
    rectangle, draw=black, thick, fill=cyan, fill opacity=0.15, text opacity=1, rounded corners
    ] at (\tinf,\ylabel){supercooled expansion begins};
\node[above,align=center,text width=1.4cm, minimum height=2em,inner sep=2pt,
    rectangle, draw=black, thick, fill=orange, fill opacity=0.15, text opacity=1, rounded corners
    ] at (\tnuc,\ylabel){bubble nucleation};
\node[above,align=center,text width=1.3cm, minimum height=2em,inner sep=2pt,
    rectangle, draw=black, thick, fill=pink, fill opacity=0.15, text opacity=1, rounded corners
    ] at (\trh,\ylabel){reheating completes};
\end{scriptsize}
\draw[thick,gray,->] (\tinf,8.75) -- (\tinf,\ylabel);
\draw[thick,gray,->] (\tnuc,8.75) -- (\tnuc,\ylabel);
\draw[thick,gray,->] (\trh,8.75) -- (\trh,\ylabel);
\node[inner sep=3pt] at (10,\ytitle){\textit{Impact of supercooling and reheating on:}};
\draw[thick,->] (3,\yFI) -- (19,\yFI);
\draw[line width=6pt,opacity=\opa,NavyBlue] (\tinf,\yFI) -- (\trh,\yFI);
\draw[line width=6pt,opacity=\opb,NavyBlue] (\trh,\yFI) -- (18.7,\yFI);
\node[left,inner sep=3pt,NavyBlue] at (3,\yFI){Freeze-in};
\draw[thick,->] (3,\yFO) -- (19,\yFO);
\draw[line width=6pt,opacity=\opa,ForestGreen] (\tinf,\yFO) -- (\tnuc,\yFO);
\draw[line width=6pt,opacity=\opb,ForestGreen] (3.3,\yFO) -- (\tinf,\yFO);
\node[left,inner sep=3pt,ForestGreen] at (3,\yFO){Freeze-out};
\draw[thick,->] (3,\yGW) -- (19,\yGW);
\draw[line width=6pt,opacity=\opa,BrickRed] (\tnuc,\yGW) -- (\trh,\yGW);
\draw[line width=6pt,opacity=\opb,BrickRed] (\tinf,\yGW) -- (\tnuc,\yGW);
\node[left,inner sep=3pt,BrickRed] at (3,\yGW){GW production};
\end{scope}
\end{tikzpicture}
\caption{Schematic evolution of the energy density in the Universe around a supercooled phase transition, and impact on the production of dark matter in a separate sector.
In this paper, we focus on the implications for direct detection of dark matter candidates whose abundance is set around the supercooled PT via freeze-out or freeze-in. 
We show how a supercooling stage can enhance the chances of detection at direct searches, and we conclude by discussing the gravitational wave background produced at the end of supercooling, and the signature of the reheating stage on the low-frequency causality tail.}
\label{fig:thermal_history}  
\end{figure}
As sketched in Fig.~\ref{fig:thermal_history}, a BSM PT could display a prolonged stage when the Universe remains trapped in the metastable vacuum, such that the vacuum energy of this configuration dominates the energy budget and determines a phase of inflation \cite{Randall:2006py, Baratella:2018pxi,Levi:2022bzt} (in addition to the commonly assumed primordial inflation stage \cite{Riotto:2002yw,Baumann:2009ds,Senatore:2016aui}).
As the universe cools down, the PT becomes energetically favourable and completes, via the nucleation at $\Tnuc$ of bubbles of true vacuum that collide in a Universe partially depleted of the thermal plasma by the SC.
Importantly, the gravitational wave background (GWB) produced in the final phase via bubble collisions is a key probe of this scenario \cite{Ellis:2019oqb,Ellis:2020nnr,Lewicki:2021xku,Gerlach:2025fkr}.
As the PT completes, the Universe is reheated to a final temperature $\TRH$. 
While the details of this stage are model dependent, a generic feature is that the EOS parameter during reheating deviates from $w\neq \tfrac 13$ and gets closer to $w=0$.
The conversion of the vacuum energy density into relativistic energy of the thermal bath increases the entropy density of the Universe, diluting the abundance of all the relic species that were present before the SC.

These two effects (the non-trivial time dependence of the equation of state and Hubble rate across the supercooling, and the final entropy injection) can significantly affect a dark sector, that we assume to be in thermal equilibrium with the bath before the SC and remains decoupled from it throughout the SC and preheating stage.
If the DM candidate achieves its final abundance (via freeze-out or freeze-in) during the SC or reheating phase, the interplay of the production and Hubble rates changes in a non-trivial way. 
Secondly, after reheating is completed, the energy density of the DM relative to the thermal bath is diluted by the entropy injection.

In this paper, we consider in detail the impact of these SC dynamics in the `visible' sector on those of the dark sector, and in particular on the freeze-out (FO) and freeze-in (FI) dynamics of the DM candidate, with an eye on the prospects for direct DM detection. 
Similar considerations about the impact of an entropy injection after SC were made in the context of indirect searches in \cite{Davoudiasl:2015vba}.

For the FO scenario, we focus on two main options.
A possible motivation for a SC stage is that the electroweak (EW) sector might be modified into a quasi-conformal theory, which undergoes a SC PT until when the EW symmetry is spontaneously broken by the QCD PT \cite{vonHarling:2017yew,Baratella:2018pxi}. 
This suggestive scenario would imply a long phase of SC (perhaps at the cost of some tuning \cite{Mishra:2024ehr, Agrawal:2025xul}) and would be linked to defined inflationary scales $\Tinf\sim \TeV$, $\Tnuc\lesssim \GeV$ \cite{Baratella:2018pxi}.
An alternative option is to assume more generic conditions for the SC, which would likely imply a shorter duration (a few $e$-folds). 
An interesting energy range where the impact on the DM dynamics might be relevant is around the FO of a WIMP-like particle, which is a motivated DM candidate. 
We discuss these two scenarios in \cref{sec:SC EWSB,sec:WIMP}, and highlight when these effect can bring some motivated WIMP-like DM candidates closer to detectability.

Another simple production mechanism for DM is FI, which has a natural realisation in a millicharged particle undergoing an IR-dominated FI around $T\sim m_\DM$ \cite{Hall:2009bx,Bernal:2017kxu,Chang:2019xva}. 
As we show in \cref{sec:supercool_freezein}, if the SC occurs around or after the FI epoch, larger couplings between DM and SM are suggested, moving the favoured parameter space (for $m_\DM \gtrsim \TeV$) closer to  reach for direct detection.

We then discuss in \cref{sec:GW} an almost guaranteed signal of a SC scenario, that is a GWB produced by the bubble collisions at the end of the inflation phase.
We highlight in particular the features that would confirm the assumptions of some of the results derived in the previous sections, i.e.~a non-instantaneous reheating phase which determines a hierarchy $\Trh < \Tinf$.
The cosmological phase of non-standard EOS parameter ($w\gtrsim 0$) imprints a peculiar feature on the low-frequency part of the GW spectrum, the so-called causality tail \cite{Watanabe:2006qe,
Caprini:2009fx,
Barenboim:2016mjm,
Saikawa:2018rcs,
Cui:2018rwi,
DEramo:2019tit,
Figueroa:2019paj,
Gouttenoire:2019kij,
Cai:2019cdl,
Hook:2020phx}. 
Notably, the longer matter-dominated phase determines a larger redshift from $\Tinf$ until today, which shifts the physical GW frequencies to lower values, effectively enhancing the amplitude of the GWB at low frequencies.
\cref{sec:conclusions} contains our conclusions.
In the appendices, we elaborate on two aspects of our study. Appendix~\ref{app:darkQED} makes the point that, if the dark sector is very weakly coupled to the SM, we do not need to worry that the DM production after the SC might overcome the relic abundance that we compute.
In Appendix~\ref{app:baryogenesis} we comment on the fate of the baryon asymmetry after the SC, depending on when and how much was generated, and elaborate on the possibility of having an asymmetric DM relic.

\section{Supercooled phase transition in the early universe}
\label{sec:SC_history}

Supercooled phase transitions occur when, due to a suppressed nucleation rate $\Gamma$, the system remains trapped in the metastable phase for extended periods, beyond the critical temperature at which a true vacuum develops. When the universe cools down and reaches the temperature $\Tinf$ defined by the equality
\begin{equation}
\label{eq:Hinf}
    \frac{\pi^2}{30}g_{*}\Tinf^4=\Delta{V} = 3 H_{\text{inf}}^2 \, m_{\text{Pl}}^2\,\,,
\end{equation}
the vacuum energy becomes the dominant component. From this point, the universe undergoes a period of inflation characterized by a constant Hubble rate $H_{\text{inf}}$ and hence an exponential expansion  
\begin{equation}
\frac{a(t)}{a(t_{\text{inf}})}=e^{H_{\text{inf}}\,t}\,,\quad \frac{T(t)}{T(t_{\text{inf}})}=e^{-H_{\text{inf}}\,t}\,.
\end{equation}
The exponential decrease of the temperature (equivalently, growth of the scale factor) can also be parametrized in terms of a number of $e$-folds $\Delta N$, defined by the relation $T=\Tinf\, e^{-\Delta N}$. The vacuum domination epoch lasts until when the universe reaches the nucleation temperature $\Tnuc$ such that the nucleation rate per volume is $\Gamma(\Tnuc) > H^4$, triggering enough bubbles formation and the completion of the transition. The subsequent release of latent heat then reheats the universe, restoring thermal equilibrium. This cosmological history is sketched in \cref{fig:thermal_history}.

Prior to supercooling, the number density of any particle species in thermal
equilibrium with the radiation bath follows the standard thermal distribution:
\begin{equation}
    n_{i}^{\textrm{eq}}=\frac{\zeta(3)}{\pi^2} g_{i} T^3\,,
\end{equation}
and the total entropy density is given by
\begin{equation}
    s=\frac{2\pi^2}{45}\gs T^3\,.
\end{equation}
Therefore, the abundance of any particle species at equilibrium reads:
\begin{equation}
Y_i^{\text{eq}}=\frac{n_{i}^{\textrm{eq}}}{s}=\frac{45 \zeta(3)g_{i} }{2 \pi^4 \gs }\,.
\end{equation}
During the phase of inflation, both the number density and entropy density scale as $a^{-3}$, ensuring that the yield $Y_i$ remains constant despite the exponential expansion of the Universe and the corresponding decrease in temperature. This applies also to relic abundance of decoupled species. On the other hand, the completion of the transition and the subsequent reheating lead to a significant entropy injection. Any pre-existing particle abundance is then diluted by the factor \cite{Davoudiasl:2015vba,Cohen:2008nb,Baldes:2020kam} 
\begin{equation}\label{eq:SCfactor}
    Y_i^{\textrm{SC}}=Y_i \cdot \frac{\Tnuc^3 \Trh}{\Tinf^4} \,.
\end{equation}
Here, $\Trh$ denotes the reheating temperature. By simple energy budget considerations, $\Trh$ equals $\Tinf$ in the case of instantaneous reheating. 
In a realistic setting, the finite-duration dynamics of the PT (latent heat release, bubble collisions, and particle acquiring a mass) determine an EOS parameter $w<\tfrac 13$ during the reheating phase, until the restoration of thermal equilibrium in the bath.

Particles may be thermally regenerated in the radiation bath after reheating, provided that $\TRH \gtrsim m_i$.\footnote{The reheating temperature $\Trh$ should be compared to the characteristic temperature at which particles normally decouple in a radiation-dominated universe. In the DM case, for example, it would be the freeze-out temperature $T_{\text{f.o.}}$ or the freeze-in temperature $T_{\text{f.i.}}$, depending on the specific model dynamics.} Consequently, the total abundance receives two contributions, schematically \cite{Hambye:2018qjv},
\begin{equation}\label{eq:totalY}
    Y^{\textrm{tot}}_i \simeq Y_i^{\textrm{SC}}+Y_i^{\textrm{th}},
\end{equation}
where $Y^{\textrm{th}}_i$ is the sub-thermal contribution posterior to reheating, and $Y_{i}^{\text{SC}}$ is the diluted particle abundance according to \cref{eq:SCfactor}. Depending on $\Trh$, one of these contributions prevails above the other. In the following, we will focus on the case in which reheating is not very efficient, so that we can neglect $Y_i^{\textrm{SC}}\ll Y_i^{\textrm{th}}$.
We discuss this condition more in detail in \cref{sec:reheating}.

This class of PTs can produce some especially strong GW signals, due to the large values of the strength parameter $\alpha =\Delta V/\rho_{\text{rad}}|_{T=\Tnuc}$ \cite{Athron:2023xlk}. A discussion of the GW spectra produced in the scenarios considered in this paper will be presented in sec.~\ref{sec:GW}.

\section{Dark matter relic in a supercooled universe}
\label{sec:DMsupercooled}
Supercooling epochs can potentially span significant portions of the Universe's thermal history, up to $\mathcal{O}(10)$ e-folds, without requiring much tuning \cite{Levi:2022bzt}. Therefore, the dilution factor of \cref{eq:SCfactor} can significantly suppress the thermal relic abundance of dark matter. As a result, the viable parameter space for mass and coupling values, such as those considered in the usual WIMP scenarios, can be drastically modified. It is therefore natural to ask, on general grounds, what impact such a non-standard cosmological epoch could have on DM, whether it be a minimal candidate or a light particle within a more complex dark sector. One should consider the possibility that the dynamical production of the DM relic occurs during any of the stages illustrated in \cref{fig:thermal_history}, and carefully account for the role of the reheating dynamics and entropy injection.

The phenomenological consequences of these non-standard scenarios on the DM thermal relic have already been investigated in a number of previous works.
For example, \cite{Davoudiasl:2015vba, Cohen:2008nb, Baratella:2018pxi} explore the scenario where freeze-out occurs before supercooling and the relic is subsequently diluted by injection of entropy as in \cref{eq:SCfactor} (a set-up dubbed \textit{Inflatable Dark Matter} in~\cite{Davoudiasl:2015vba}), 
and \cite{Hambye:2018qjv, Baratella:2018pxi, Baldes:2020kam,Kierkla:2022odc,Rescigno:2025ong}(and references therein)  consider the case where initially massless particles acquire a mass through the phase transition, forming elementary or composite states that can be heavier or lighter than the reheating temperature (nicknamed \textit{Supercool Dark Matter} in \cite{Hambye:2018qjv})%
\footnote{First-order PTs have also been proposed as production mechanisms for primordial black holes, see e.g.~\cite{Gouttenoire:2023naa,Salvio:2023ynn,Banerjee:2024cwv,Kierkla:2025vwp, Franciolini:2025ztf} and references therein.}. 
In the following sections, we discuss the possibility of DM relic production during the vacuum domination period, region~II in \cref{fig:thermal_history}, where the standard picture of particle decoupling (see e.g.~\cite{Kolb:1990vq}) is modified. During this phase, the Hubble rate no longer scales as $T^2$, but instead remains approximately constant, $H \simeq H_{\text{inf}}$, as defined in \cref{eq:Hinf}. As a result, for a given temperature dependence of $\Gamma_{\text{int}}$, the decoupling condition $\Gamma_{\text{int}} \approx H$ can be satisfied at a different temperature compared to the standard radiation-dominated case. This shift in the decoupling point ultimately impacts the resulting relic abundance. 

While the possibility of producing DM through the non-equilibrium dynamics of the first-order PT is certainly appealing\footnote{In a first-order PT featuring runaway bubbles, heavy DM particles can be abundantly produced either during bubble collisions \cite{Konstandin:2011ds, Falkowski:2012fb, Katz:2016adq,Shakya:2023kjf,Mansour:2023fwj, Cataldi:2024pgt, Giudice:2024tcp} or through interactions between the bubble walls and the surrounding plasma \cite{Azatov:2020ufh,  Azatov:2021irb,Azatov:2021ifm, Azatov:2024crd}. Furthermore, particle production may also occur during the tachyonic or preheating stages associated with the oscillations of the scalar field \cite{Felder:2001kt,Schmitt:2024pby, Girmohanta:2025wcq}.}, we focus here on scenarios in which the dark sector can be independent of the new physics responsible for the transition. Although that possibility may offer a minimal solution, it is still interesting to explore the phenomenology of a dark sector under general assumptions\footnote{As pointed out in \cite{Baldes:2020kam}, achieving the correct relic abundance of DM composite particles arising from a supercooled confining transition requires a prolonged period of SC, up to $\Delta N \approx 13$, which is argued to be tuned in recent studies on strongly coupled dynamics \cite{Mishra:2024ehr,Agrawal:2025wvf}.}.

We now present a quantitative analysis of the impact of supercooling on the freeze-out and freeze-in dynamics of dark matter. Along with discussing these effects in general terms, we illustrate our findings with applications to motivated benchmark models.

\subsection{Dark matter freeze-out completing during supercooling}\label{sec:supercool_freezeout}

In a standard radiation-dominated universe the Boltzmann equation can be written in the well-known compact form \cite{Kolb:1990vq}:
\begin{equation}
\frac{dY}{dx} = \frac{\lambda}{x^2}\left(Y^2 - Y_{\text{eq}}^2\right), \quad \text{where} \quad \lambda = m_{\text{Pl}} m_{\DM} \langle \sigma_{\text{ann}} v_{\text{rel}} \rangle \sqrt{\frac{8 \pi^2 \gs }{45}}\,,
\end{equation}
where $m_{\text{Pl}}$ denotes the reduced Planck mass and assuming $g_{\star}=\gs$. The $1/x^2$ dependence on the right-hand side reflects the fact that the Hubble rate decreases quadratically with temperature, i.e., $H \sim T^2$. However, if the cosmological history of the universe includes a period of supercooling, the form of the Boltzmann equation is modified. In particular, if freeze-out occurs during the phase of inflation the resulting dynamics differ substantially. As the Hubble parameter becomes constant from the time when the radiation and vacuum energy densities are equal, the Boltzmann equation takes the form:
\begin{equation} \label{eq:BoltzmannInf}
\frac{dY}{dx} = \frac{\mu}{x^4}(Y^2 - Y_{\text{eq}}^2)\,, \quad \text{where} \quad \mu = \frac{2 \pi^2}{45}\gs \frac{m_{\DM}^3\langle \sigma_{\text{ann}}v_{\text{rel}}\rangle}{H_{\text{inf}}} \,,
\end{equation}
with the parametric difference coming from the different Jacobian factor in $dt/dx$.
To see how this compare to the standard Boltzmann eq., it is useful to consider the ratio:
\begin{equation}
\frac{\lambda}{\mu} = \sqrt{\frac{90}{\pi^2 \gs}} \frac{m_{\text{Pl}} H_{\text{inf}}}{m_{\DM}^2} = \frac{T_{\text{inf}}^2}{m_{\DM}^2} = x^{-2} \frac{T_{\text{inf}}^2}{T^2} \simeq x^{-2} e^{2H_{\text{inf}} \Delta t}\,,
\end{equation}
where $\Delta t$ denotes the time elapsed since the onset of inflation. Thus, to a first approximation, the two equations are similar, but the effective thermally averaged annihilation cross section during supercooling is exponentially decreasing in time, relative to that in a radiation-dominated universe:
\begin{equation}\label{eq:effsigma}
\langle \sigma_{\text{ann}} v_{\text{rel}} \rangle_{\text{inf}} \sim \langle \sigma_{\text{ann}} v_{\text{rel}} \rangle \, e^{-2 H_{\text{inf}} \Delta t}.
\end{equation}
We can then expect that particles decouple earlier in an inflating universe compared to standard cosmology. 

When freeze-out occurs in the non-relativistic regime and in standard radiation dominated universe, the time at which the particle decouples from the bath scales approximately as $x_{\text{f.o.}}^{\text{rad}} \simeq \log(\lambda)$. Therefore, approximating $\mu/x^2\to \mu/(x_{\text{f.o.}}^\textsc{sc})^2$ in the supercooled case, the shift in freeze-out between the two scenarios can be crudely estimated as:
\begin{equation}
\Delta x_{\text{f.o.}} \propto \log\left(\frac{\lambda }{\mu / (x_{\text{f.o.}}^\textsc{sc})^2}\right) = 2H \Delta t_{\text{f.o.}}^\textsc{sc} = 2 \Delta N_{\text{f.o.}}^\textsc{sc}\,,
\end{equation}
\begin{figure}[t!]
\centering 
\includegraphics[width=0.75\textwidth]{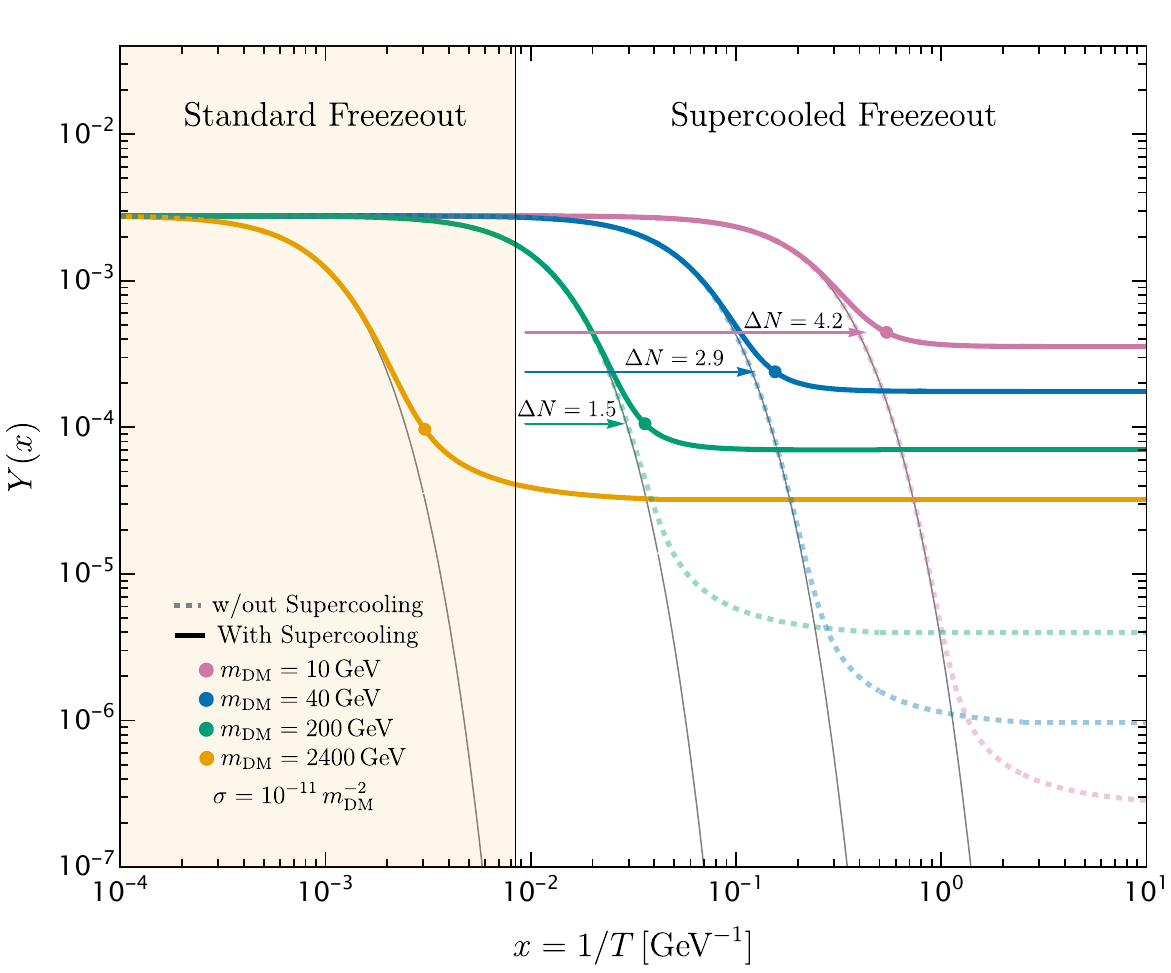}
\caption{
DM freeze-out for several masses. The plotted yield $Y$ is evaluated before the end of supercooling and the consequent entropy injection, that dilutes $Y$ by a factor $(\Tnuc/\Tinf)^3$. The solid lines are the full solution to the Boltzmann equation while the circular dots indicate the moment where $\Delta Y(x)/Y_{\text{EQ}}(x)=2$, that we identify as freeze-out condition. For comparison, we show as dashed lines the solution that one would get for same mass and cross section in a standard radiation dominated universe. The vertical line describes the onset of supercooling while $\Delta N$ is the number of $e$-folds elapsed since that moment.  In these plots $\Tinf =120\, \GeV$ and $\sigma=10^{-11}m_{\DM}^{-2}$.}
\label{fig:SC_freezeout}  
\end{figure}
where $\Delta N_{\text{f.o.}}^\textsc{sc}$ is the number of e-folds until the freeze out. This shift could be significant enough to yield a noticeable difference in the final relic abundance due to the exponential dependence of the Boltzmann distribution $n\simeq x^{3/2}\exp(- x)$.
To have a rough estimate of this effect, we can use the cold relic assumption: when freeze-out occurs in the non-relativistic regime, the final relic yield is approximately proportional to $x_{\text{f.o.}}$ and inversely proportional to the annihilation cross section. Therefore, using the previous approximations, we can derive an estimate for the ratio between the relic abundances in the supercooled scenario and the standard radiation-dominated case
\begin{equation}
\label{eq:SCYield}
\frac{Y^{\text{SC}}_{\infty}}{Y^{\text{rad}}_{\infty}}\approx\frac{x_{\text{f.o.}}^{\text{SC}}}{x_{\text{f.o.}}^{\text{rad}}} \cdot e^{2 \Delta N_{\text{f.o.}}^{\text{SC}}}
\approx \left(1+\frac{2 \Delta N_{\text{f.o.}}^{\text{SC}}}{x_{\text{f.o.}}^{\text{SC}}}\right)^{-1}e^{2 \Delta N_{\text{f.o.}}^{\text{SC}}}\,.
\end{equation}

As an illustrative example, some numerical solutions to the Boltzmann equations are presented in \cref{fig:SC_freezeout}, where we fix $\Tinf \simeq 120 \, \GeV$ and analyze the freeze-out dynamics for a range of particle masses $m_{\DM} = \mathcal{O}(10)\,\GeV$--$\mathcal{O}(1)\,\TeV$. We assume a parametric dependence of the dominant $s$-wave annihilation cross section given by $\sigma \simeq 10^{-11} m_{\DM}^{-2}$ and fix the number of degrees of freedom to $g_\star=\gs=150$. Regardless of the details of these parameter choices (for which we discuss a possible realisation in the following), we emphasize that the qualitative effects described here may be relevant across a broader class of scenarios involving higher temperature and mass scales.

The results in \cref{fig:SC_freezeout} confirm the expectation that the final yield of a species experiencing a period of supercooling lasting $\Delta N$ $e$-folds is exponentially enhanced by a factor proportional to $e^{2 \Delta N}$ with respect to the standard freeze-out scenario. Indeed, \cref{eq:SCYield} provides reasonably good predictions: $Y_{\infty}^{\text{SC}}/Y_{\infty}^{\text{rad}}\simeq 1646,176,13$ respectively for the purple, blue, and green lines. Therefore, the particle mass required to achieve the correct relic abundance differs exponentially from the value expected in a radiation-dominated universe. Remarkably, smaller masses, despite corresponding to larger annihilation cross sections for our parametrisation $\sigma\propto m_\DM^{-2}$, lead to a larger relic abundance. This can be understood by inspecting \cref{eq:effsigma}, as smaller masses imply a later freeze-out and a larger exponential suppression of the effective cross-section of the Boltzmann equation. This behavior is in stark contrast to the standard scenario in radiation domination, where a larger cross section would result in smaller relic abundances.

\subsubsection{Example: supercooled electro-weak transition}
\label{sec:SC EWSB}
We now analyze an explicit realization of these effects by considering the supercooled scenarios discussed in \cite{Randall:2006py, vonHarling:2017yew, Baratella:2018pxi}, within the framework of strongly coupled gauge extensions of the SM motivated, for instance, by compositeness. 
Studies of these classes of theories and their five-dimensional duals based on the AdS/CFT correspondence, show that the transition from the deconfined to the confined phase can be a first-order PT accompanied by a period of SC \cite{%
Randall:2006py,
Nardini:2007me,
Konstandin:2010cd,
Konstandin:2011ds,
Konstandin:2011dr,
vonHarling:2017yew,
Megias:2018sxv,
Baratella:2018pxi,
Mishra:2024ehr,
Agrawal:2025wvf}.  Whether and under what conditions such SC can occur remains the subject of increasingly refined investigations \cite{Agrawal:2025wvf, Gherghetta:2025krk}.

These scenarios can be qualitatively summarised as follows. Although the confined phase becomes energetically favourable below a certain critical temperature, the tunnelling rate between the two phases can be highly suppressed, causing the theory to remain trapped in the deconfined phase. As the universe continues to cool and undergoes supercooling, the temperature eventually drops to the QCD scale, where QCD itself confines and begins to influence the dynamics of the extended sector. This interplay is expected on general grounds, as realistic models of compositeness tipically feature couplings between SM quarks and operators of the TeV-scale strong sector, interactions necessary for generating fermion masses after electroweak symmetry breaking. Therefore, QCD modifies the vacuum structure and triggers the completion of the phase transition.

We refer the reader to \cite{Chacko:2012sy, vonHarling:2017yew, Baratella:2018pxi} for detailed discussions on the treatment of these theories and the characterization of their phases. For practical purposes, we assume that the critical temperature marking the onset of the inflation epoch is approximately $\Tinf \sim \mathcal{O}(100)$~GeV, while bubble nucleation is triggered at temperatures around $\Tnuc \sim \LQCD^{\text{dec}}\simeq O(1)\,\LQCD$. Here, $\LQCD^{\text{dec}}$ denotes the QCD scale in the deconfined phase, which generally differs from its value in the SM vacuum, $\LQCD$, since the conformal sector, containing states charged under QCD, can modify the running of the strong coupling $\alpha_s$. SC periods spanning from the TeV scale down to the QCD scale can also arise in weakly-coupled extensions of the SM~\cite{Hambye:2018qjv, Sagunski:2023ynd, Schmitt:2024pby,Zhang:2025kbu}.

Let us now assume that the SM is extended by new degrees of freedom that give rise to a supercooled PT, as in the scenarios discussed above. In addition to this new dynamics, we consider the presence of a dark sector that remains in thermal equilibrium with the visible sector throughout the evolution of the universe, but whose impact on the spontaneous breaking of conformal invariance is negligible. In detail, we consider a simple $Z'$ model, where the vector boson is heavier than the fermionic DM candidate $\chi$ ($M_{Z^{\prime}} > 2 m_{\DM}$). The annihilation cross section in the case of vector couplings then reads \cite{Abdallah:2015ter}\footnote{For simplicity, we do not include the possible contribution from BSM fermions charged under $U(1)'$, which would not alter significantly the results.}:
\begin{equation}\label{eq:annihilation}
    (\sigma v)(\chi \bar{\chi} \rightarrow Z^{\prime} \rightarrow q \bar{q}) \approx \frac{3 m_{\DM}^{2}}{ \pi\left[\left(M_{Z^{\prime}}^{2}-4 m_{\DM}^{2} \right)^{2} + \Gamma_{Z^{\prime}}^{2} M_{Z^{\prime}}^{2}\right]} \cdot \left(g_{\chi}^{V}\right)^{2} \left(g_{q}^{V}\right)^{2}\,.
\end{equation}
This cross section enters the Boltzmann equation and governs the decoupling of DM particles in the early universe.
If the vector mediator has a mass significantly larger than the DM particle, e.g.\ $M_{Z^{\prime}} \simeq \mathcal{O}(100)\,m_{\DM}$, and we fix the ratio $M_{Z'}/m_\DM$, the annihilation cross section exhibits a parametric dependence $\sigma \propto m_\DM^{-2}$.
\begin{figure}[t!]
\centering
\includegraphics[width=0.75\textwidth]{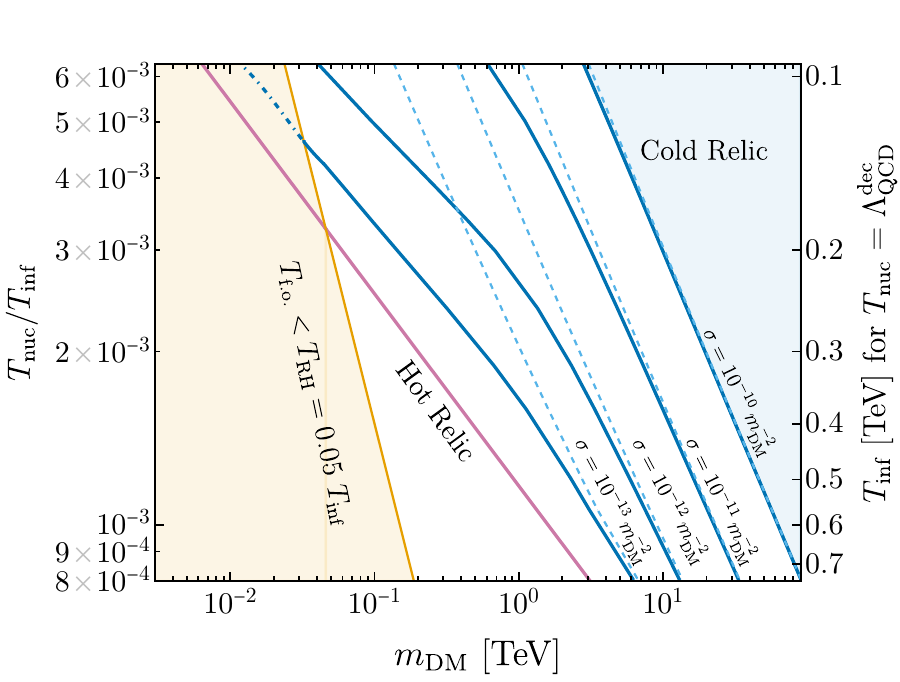} 
\caption{
The solid blue lines show the required supercooling as a function of DM masses for several annihilation cross sections $\sigma$, ranging from $\sigma=10^{-13} m_{\DM}^{-2}$ to $\sigma=10^{-10} m_{\DM}^{-2}$. We fix here $\TRH=0.05\,\Tinf$. The dashed lines represent the results that we would obtain by considering a radiation-dominated universe. In purple, we show the analytic approximation of \cref{eq:hotRelic}.}
\label{fig:DMrelic}  
\end{figure}
We show in \cref{fig:DMrelic} the region of parameter space in the $(m_{\DM}$ vs. $\Tnuc/\Tinf)$ plane that is compatible with the observed dark matter relic abundance, for different values of the annihilation cross section. For concreteness, we fix $\Trh = 0.05 \,\Tinf$, which is indicated by the shaded orange region in the plot.

For heavy DM candidates, \textit{i.e.}~in the cold relic regime shown in \cref{fig:DMrelic}, freeze-out occurs prior to the onset of SC. In this case, the final relic abundance can be estimated by
\begin{equation}\label{eq:coldRelicSC}
    \Omega_{\text{DM}}^{\text{cold}} h^2 \simeq 1.07 \cdot 10^9 \frac{(n+1)x_{\text{f.o.}}^{n+1}\,\text{GeV}^{-1}}{(\gs / g_{*}^{1/2}) M_{\text{Pl}}\, \sigma} \cdot \left(\frac{\Tnuc}{\Tinf} \right)^3 \cdot \left(\frac{\Trh}{\Tinf} \right)\,.
\end{equation}
Freeze-out may also occur before the supercooling phase in the case of a very small annihilation cross section, resulting in a hot relic, that we represent as a purple curve in \cref{fig:DMrelic}. This scenario is described by
\begin{equation}\label{eq:hotRelic}
    \Omega_{\text{DM}}^{\text{hot}} h^2 \simeq 0.075 \cdot \left(\frac{m}{\eV}\right) \cdot \left(\frac{g_\DM}{\gs (x_f)} \right) \cdot \left(\frac{\Tnuc}{\Tinf} \right)^3 \cdot \left(\frac{\Trh}{\Tinf} \right)\,,
\end{equation}
where $g_\DM$ is the number of internal degrees of freedom of the DM particle. Besides these two limiting cases, we observe that for DM candidates with masses below $m_{\DM} \sim \mathcal{O}(1)\,\TeV$, and thus freeze-out temperatures $\Tfo$ expected within the SC window, a detailed numerical solution of the Boltzmann equation is required. The numerical results (solid lines) can indeed differ by an order of magnitude or more compared to including just the dilution factor (dashed lines), shifting the viable mass range. In general, smaller masses lead to freeze-out occurring earlier compared to standard cosmology (see also \cref{fig:SC_freezeout}), and so closer to the relativistic regime where the Boltzmann equilibrium distribution $Y(x)$ flattens at small $x$. As a result, the leftmost blue curves in \cref{fig:DMrelic} resemble the behaviour typical of hot DM. We may then conclude that supercooled DM, when decoupling during the supercooling phase, displays a freeze-out closer to a warmer relic, compared to the standard freeze-out.

As discussed above, if the reheating temperature is large enough such that $\Trh > \Tfo$, DM particles can be thermally regenerated and potentially reach equilibrium again. This limit is stressed by the dot-dashed curves in \cref{fig:DMrelic}. In such scenarios, reheating effectively erases any memory of the supercooling phase, and freeze-out proceeds through the standard dynamics of a radiation-dominated universe. Therefore, the characteristic enhancement of the DM relic abundance due to supercooling is only relevant within the temperature window $\Tinf - \Trh$. The size of this window depends on the model-dependent reheating dynamics. We discuss more details about this point in \cref{sec:reheating}.

\begin{figure}[t!]
\centering
\includegraphics[width=0.85\textwidth]{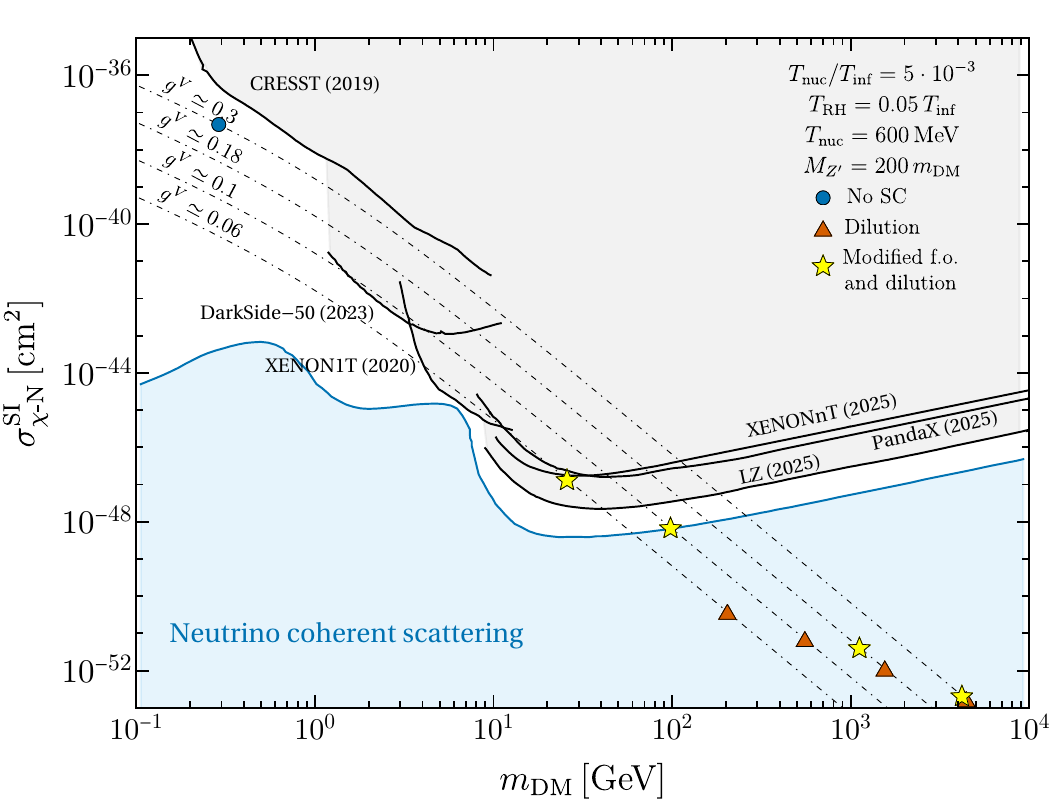} 
\caption{
Constraints from direct detection experiments on the spin-independent dark matter–nucleon cross section as a function of the dark matter mass, including results from LZ~\cite{LZ:2024zvo}, XENONnT~\cite{XENON:2025vwd}, PandaX~\cite{PandaX:2024qfu}, DarkSide-50 \cite{DarkSide-50:2022qzh}, and CRESST \cite{CRESST:2019jnq}. The red triangles and yellow stars indicate the points that yield the correct relic abundance considering, respectively, just the naive dilution factor or solving the full numerical freeze-out during supercooling. The blue marker indicates the dark matter mass required in the absence of supercooling for this model. Neutrino fog defined according to \cite{OHare:2021utq}.}
\label{fig:DirectSearch}  
\end{figure}

If we fix its couplings to SM fermions, the DM mass $m_{\DM}$ can be determined by requiring the correct relic abundance. 
It is then possible to compute the spin-independent DM–nucleon cross section relevant for direct detection experiments:
\begin{equation}
    \sigma_{\chi-N}^{\mathrm{SI}} = 1.13 \times 10^{-43}~\mathrm{cm}^{2} \cdot (g_{\chi}^{V} g_{q}^{V})^2 \left(\frac{\mu_{\chi-N}}{1\,\GeV}\right)^{2} \left(\frac{10\,\TeV}{M_{Z^{\prime}}}\right)^{4},
\end{equation}
where $\mu_{\chi-N}$ is the reduced mass of the DM–nucleon system. Results for several choices of the coupling $g_{\chi}^{V} = g_{q}^{V} \equiv g^V$ are shown in \cref{fig:DirectSearch}. 
The orange triangle and the yellow star correspond to the parameter values matching the observed relic abundance when considering, respectively, the naive dilution factor and the full numerical solution of freeze-out during SC.
The case of a standard radiation-dominated cosmology is marked with a blue dot in \cref{fig:DirectSearch} and, for these coupling choices, leads to $m_{\DM}\leq 0.1\,\GeV$, which can be probed via electron scattering \cite{Essig:2011nj,Graham:2012su, Essig:2015cda, Hochberg:2015pha, Essig:2022dfa,
QROCODILE:2024nqm, Griffin:2024cew}. 

The effect discussed in \cref{sec:supercool_freezeout} is very significant in the context of direct detection. 
Since the spin-independent nuclear scattering cross section scales as $M_{Z'}^{-4} \propto m_\DM^{-4}$ (for $m_\DM > \GeV$ so that $\mu_{\chi-N}$ is constant, and fixing $M_{Z'}/m_\DM$), a mismatch of just one order of magnitude in the predicted $m_\DM$ can translate into a four-order-of-magnitude deviation in the expected nuclear cross section. Parametrically, the correction to the cross section is enhanced by a factor roughly proportional to $e^{8 \Delta N}$.

If one estimates the relic abundance using only the naive entropy dilution factor, the resulting DM mass tends to be relatively large, which in turn yields overly suppressed nuclear cross sections. These predictions often fall deep into the so-called neutrino fog, where coherent neutrino scattering obscures DM signals. 
In contrast, when the freeze-out process is treated accurately by solving the Boltzmann equation within the proper supercooled cosmological background, the predicted $m_\DM$ can be significantly lower. 
This leads to cross sections that can lie within the reach of current or near-future direct detection experiments. 

\subsubsection{Example: supercooling during the freeze-out of a WIMP candidate}
\label{sec:WIMP}
The results displayed in Fig.~\ref{fig:DirectSearch} illustrate how the non-standard cosmologies discussed in this work can shift the predicted DM target mass on the $\sigma^{\text{SI}}_{\chi-N}$ plane, depending on the supercooling window. 
So far, we focused on a supercooled PT taking place roughly between the EW and QCD scales, and we showed the implications for DM masses and couplings somewhat smaller than the standard WIMP candidate.
It is therefore interesting to examine the potential impact of this effect on motivated WIMP-like particles, which constitute a primary target for  direct detection experiments. 
Since these candidates (for the case of spin-independent cross sections that we consider here) often occupy regions of parameter space that are already constrained or will be probed \cite{XENON:2025vwd, LZ:2024zvo, PandaX:2024qfu}, a long SC period accompanied by a large dilution factor (such as that considered in the previous section), would suppress the DM abundance too much.
\begin{figure}[t!]
\centering
\includegraphics[width=0.8\textwidth]{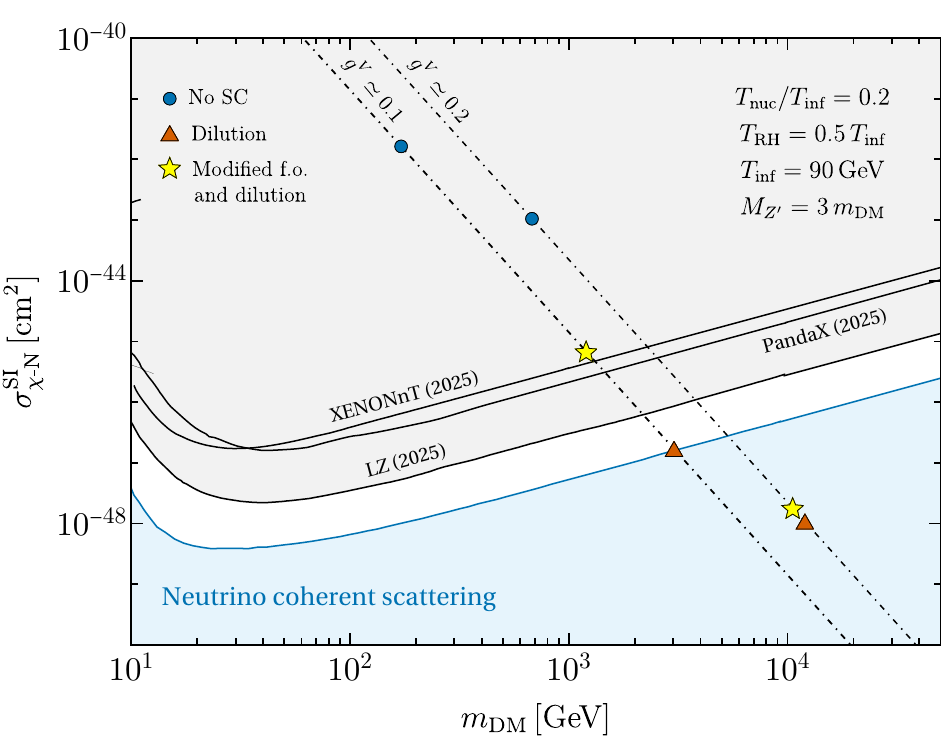}
\caption{
Same as \cref{fig:DirectSearch}, considering now DM candidates with masses and couplings fulfilling the standard ``WIMP miracle'' and a short period of supercooling.
}
\label{fig:DirectSearchWIMP}
\end{figure}
Therefore, it is instructive to explore how a shorter period of supercooling above the EW scale could affect the predictions for WIMP-like candidates. 
Such a scenario is illustrated in \cref{fig:DirectSearchWIMP}, where we present the direct detection constraints, analogous to \cref{fig:DirectSearch}, for a different SC window and larger cross sections. This is achieved by assuming that the $Z'$ mediator is not significantly heavier than DM.

In this example, DM candidates that would not be viable under the standard thermal history can become viable and/or move into a more interesting region of the parameter space once we account for the effects of the dilution after SC, and the modifications to freeze-out that we illustrated. 
Again, these results highlight the importance of a precise treatment of the freeze-out process in cosmologies featuring non-standard thermal histories~\cite{Arcadi:2025ifl}.

\subsection{Dark matter freeze-in during supercooling}\label{sec:supercool_freezein}

The modified thermal history of SC can also affect the freeze-in mechanisms. As before, the main consequence is the overall suppression of the DM relic abundance produced before the end of the supercooling epoch, caused by the large entropy injection and the resulting dilution factor $\left(\Tnuc/\Tinf\right)^3\left(\Trh/\Tinf\right)$. In addition, and similarly to the freeze-out case, the altered thermal history also leads to a modified Boltzmann equation, which in turn affects the amount of DM produced via freeze-in within the inflationary window. 

In standard cosmology, the yield $Y$ typically grows monotonically as 
\begin{equation}
    \frac{dY}{dT} \sim -\frac{1}{T^{2}}\,,
\end{equation}
until it reaches a plateau at $T_{\text{f.i.}}^{\textrm{rad}}\simeq \max(m_{\DM}, m_{\SM})$, where $m_{\SM}$ denotes the mass of the SM species responsible for producing DM particles \cite{Hall:2009bx}. However, if $m_\DM$ lies within the supercooling regime, the yield $Y$ no longer grows as $1/T$, due to the approximately constant Hubble rate. Instead, it reaches its freeze-in value at a temperature around $\Tinf > T_{\text{f.i.}}^{\textrm{rad}}$, resulting in a reduced production of DM particles.
Remarkably, SC affects freeze-in in the opposite way compared to freeze-out, leading to a suppression rather than an enhancement of the final relic abundance.

\begin{figure}[t!]
\centering
\includegraphics[width=0.8\textwidth]{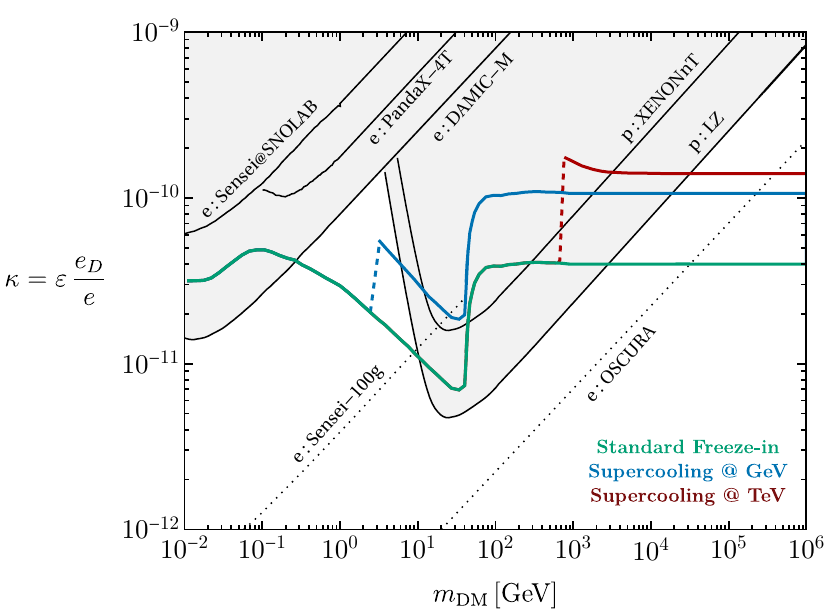} 
\caption{Predictions for the dark coupling and mixing parameters in the presence of a supercooling epoch, respectively for $\Tnuc^3 \Trh/\Tinf^4=0.14(0.08)$ for the blue (red) line. The green band indicates the standard freeze-in result. The dashed blue and red lines emphasize that predictions around the reheating transient are model-dependent and beyond the scope of this work. The gray regions are excluded by Sensei@SNOLAB~\cite{SENSEI:2023zdf}, PandaX-4T~\cite{PandaX:2022xqx}, DAMIC-M~\cite{DAMIC-M:2025luv}, LZ~\cite{LZ:2024zvo}, XENON-nT~\cite{XENON:2025vwd}.}
\label{fig:DDfreezein}  
\end{figure}

We explicitly study these effects in the context of dark QED \cite{Chu:2011be,Hambye:2019dwd,Iles:2024zka}, featuring a Dirac fermion with mass $m_\DM$ and charge $e_D$ under a dark Abelian gauge group $U(1)_D$. In \cref{fig:DDfreezein} we show the predicted values of $m_{\DM}$ and the coupling $\kappa \equiv \varepsilon\, e_D / e$ that reproduce the correct relic abundance. Here, $\varepsilon$ denotes the kinetic mixing parameter between the dark and visible photon, and $e$ is the SM QED coupling \cite{Chu:2011be}.

The green solid line shows the prediction obtained in a radiation-dominated universe. The blue and red lines correspond to the predictions for a thermal history featuring a supercooled first-order PT occurring around the \GeV\ or \TeV\ scales. The dashed line indicates that the dynamics near the reheating regime are model-dependent, as in those regions DM is appreciably produced during the reheating phase; however, on general grounds, we expect the curve to smoothly reconnect to the standard freeze-in prediction for production taking place at lower temperatures.

Since the predicted value of $\kappa$ increases with the duration of the inflationary phase, we focus on short supercooling epochs in order to avoid overly stringent bounds from direct detection experiments. The temperature range in which a full solution of the modified Boltzmann equation is required is therefore reduced compared, for example, to the case shown in \cref{fig:DirectSearch}. Nevertheless, this effect can still be seen in the mildly increasing values of $\kappa$ in correspondence with the onset of the supercooling period.

A few remarks are in order concerning the direct detection limits obtained by LZ~\cite{LZ:2024zvo} and XENON-nT~\cite{XENON:2025vwd}. Experimental bounds on DM--nucleon interactions are typically derived under the assumption that scattering is mediated by a heavy particle. Although the differential event rates for a fixed DM mass generally differ in shape (owing to the different form factor $F= 1/q^2$ rather than $F=1$) between heavy- and light-mediator scenarios, Ref.~\cite{Hambye:2018qjv} showed that, at least within a given recoil energy range,  they can resemble each other for appropriately chosen DM masses. As a consequence, bounds derived under the heavy-mediator assumption can be recast as constraints on milli-charged DM by requiring the two differential event rates to match. 
Ref.~\cite{Hambye:2018qjv} explicitly performs such an analysis using XENON-1T data~\cite{XENON:2018voc}. We do not attempt to repeat a similar study here, as it lies beyond the scope of this work. We assume that an improvement in the limits on DM--nucleon scattering in the short-range (heavy-mediator) scenario can be translated into an improvement by the same factor in the long-range interaction case (mediated by the dark photon). Starting from the latest results by LZ~\cite{LZ:2024zvo} and XENON-nT~\cite{XENON:2025vwd}, we thus find the bounds illustrated in \cref{fig:DDfreezein}. See also \cite{ShamsEsHaghi:2025kci,Boddy:2024vgt} for a discussion of how modifications to the thermal history alter the predicted abundances in benchmark freeze-in models.

\subsection{Constraints from the Reheating Phase} 
\label{sec:reheating}
After the completion of the transition, a reheating transient occurs, during which the energy density $\Delta V$ is transferred to the particles in the plasma. If reheating is instantaneous and the change in the relativistic degrees of freedom $g_*$ is neglected, the final reheating temperature is simply given by $\Trh = \Tinf$. On the other hand, if the rate of energy transfer $\gamma$ is smaller than the Hubble rate $\Hinf$, the scalar field oscillates around the minimum of its potential for an extended period, corresponding to a matter-dominated phase in the expansion of the Universe. 
The reheating temperature $\Trh$ can be determined by solving the coupled system \cite{Giudice:2000ex, Hambye:2018qjv, Gonstal:2025qky}:
\begin{equation}\label{eq:reheating}
    \dot{\rho}_{\text{rad}}+4 H \rho_{\text{rad}}=\gamma\, \rho_{\phi}, \quad \dot{\rho}_{\phi}+3 H \rho_{\phi}=-\gamma\, \rho_{\phi}, \quad H^{2}=\frac{\rho_{\text{rad}}+\rho_{\phi}}{3 m_{\text{Pl}}^{2}},
\end{equation}
under the assumption that the scalar $\phi$ can be treated as an homogeneous oscillating field. 

As discussed in \cref{sec:SC_history}, DM particles can be thermally produced during reheating, potentially restoring equilibrium between the DM and SM sectors and thereby erasing any memory of the decoupling dynamics that occurred during the SC epoch.\footnote{See \cite{Benso:2025vgm} for a discussion on DM production via non renormalizable interactions dominated by the reheating period following the PT, dubbed as \textit{phase-in} by the authors.} In such a case, DM phenomenology would no longer retain any imprint of the inflationary epoch. To ensure that thermal production of DM during reheating remains negligible, the reheating temperature $\Trh$ must be sufficiently low, which requires a small ratio $\gamma / H_{\text{inf}}$ and a correspondingly prolonged matter-dominated period. Under these conditions, there exists a range of temperatures during the supercooled epoch for which the final relic abundance is determined by the freeze-out (in) dynamics described in Sec.~\ref{sec:supercool_freezeout} (\ref{sec:supercool_freezein}), and remains unaffected by the subsequent reheating process. However, we stress that the precise requirements on $\Trh$ strongly depend on the model and on the interactions between the PT scalar field and the visible and dark sectors, and should therefore be carefully analysed on a case-by-case basis. For example, the reheating of the dark sector may be governed by a feeble portal to the SM plasma, such that the DM production from the reheated visible plasma is suppressed by small coupling parameters. This occurs, for instance, in models where freeze-out takes place within the dark sector through interactions between dark photons and dark fermions, while the portal to the visible sector is suppressed by the square of the kinetic mixing parameter between the ordinary photon and the dark vector \cite{Chu:2011be}. In this case, sub-thermal DM production due to reheating can be neglected, even for large $\Trh$. We comment further on this option in Appendix~\ref{app:darkQED}.

Finally, we note that the DM abundance may consist of an asymmetric relic. If DM production during reheating is symmetric, and as long as there exists an efficient annihilation channel for this component, the constraints from the reheating period do not apply. We elaborate on this further in Appendix~\ref{app:baryogenesis}.

\section{Signatures in Gravitational Wave backgrounds}
\label{sec:GW}

Among the compelling reasons to study the phenomenology of supercooled PTs, lie the bright prospects for the detection of the associated GWB.
The exponential dilution of the thermal plasma during SC ensures that, at the completion of the PT, the released latent heat can dominate the energy budget, and the friction experienced by the bubble walls is minimised, making them reach relativistic speeds and maximising the GW production \cite{Ellis:2019oqb,Ellis:2020nnr,Lewicki:2021xku}.

For the scenarios considered in this paper, the most interesting effects arise if SC occurred roughly in the range between some hundreds of MeV and some TeVs, which correspond to a frequency range for the peak of the GWB between $\mathcal O(10)$ nHz and $\mathcal O(100)\, \mu$Hz. 
At present, the lower end of this frequency range is observed through Pulsar Timing Array measurements \cite{NANOGrav:2023gor,EPTA:2023fyk,Reardon:2023gzh,Xu:2023wog}, which have recently detected strong evidence for a GWB whose origin (astrophysical or primordial) is still to be settled \cite{Franciolini:2023wjm, Ellis:2023oxs,Babak:2024yhu,Konstandin:2025ifn,Domcke:2025esw,Lamb:2025niq}. 
A primordial origin for this signal would hint at a loud source around $\mathcal O(1-10)$ GeV, which motivates one of the cases shown in \cref{fig:DDfreezein}.
Towards larger frequencies, the LISA experiment will reach a peak sensitivity around mHz, accessing primordial sources around TeV and above.
Finally, for the intermediate frequency range, various detection schemes have been suggested \cite{Blas:2021mpc,Fedderke:2021kuy,Fedderke:2022kxq}.

\begin{figure}[t!]
\centering
\includegraphics[width=0.6\textwidth]{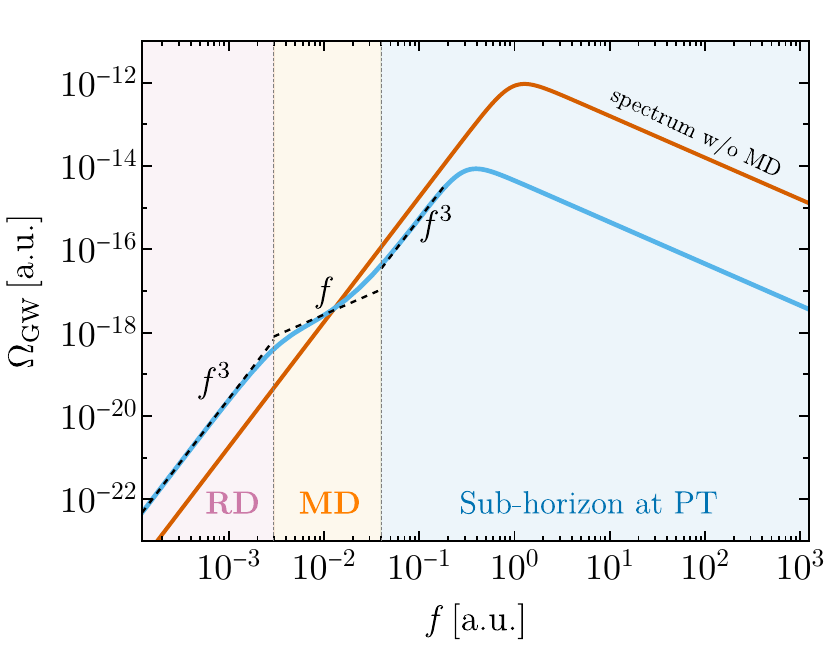} 
\caption{Example of a GW spectrum resulting from a supercooled phase transition. The orange curve shows a naive computation using the envelope approximation, assuming a radiation-dominated universe since the time of production. The blue curve instead illustrates the impact of the two effects discussed in the text: 1) the modified redshift of the energy density and of frequencies, and 2) the modified power-law scaling in the IR tail of the spectrum.}
\label{fig:GWB MD}  
\end{figure}

The study of the GW spectrum generated by primordial PTs around the peak and at larger frequencies is still an open field, with many ongoing studies \cite{Curtin:2016urg, Croon:2020cgk, Navarrete:2025yxy, Chala:2025cya, Christiansen:2025xhv} revisiting the  uncertainties in modelling the GW sources (bubble collisions, acoustic waves in the plasma, magnetic turbulence). 
For the illustrative purposes of this section, it is sufficient to fix the ideas by relying on the envelope approximation, and we model the spectrum of GWs at emission as a broken power law \cite{Kosowsky:1992vn,Huber:2008hg,Weir:2016tov}:
\begin{equation}
\label{eq:OGW emission}
\OGW h^2 = \Omega_\text{pk} \frac{(f/\fpk)^3}{1+ a (f/\fpk)^4}
\end{equation}
where $a=3$ and the peak frequency $\fpk$ and amplitude $\Omega_\text{pk}$ depend on the parameters $\alpha$, $\beta$ of the PT (see e.g.~the reviews \cite{Caprini:2018mtu, Caprini:2019egz, Athron:2023xlk}). 
In the following, we use the common notation $H_*$, $T_*$ for the Hubble rate and temperature at the nucleation time. 
The connection to the notation used in the previous sections (that is standard in the context of supercooling) is $H_*\leftrightarrow \Hinf$, $T_*\leftrightarrow \Tnuc$. 
We assume for concreteness $\beta/H_*\sim \mathcal O(10^2)$, as visible in \cref{fig:GWB MD} from the ratio of the peak frequency $\fpk$ and the frequency corresponding to the onset of MD at $H_*$.

The problem of inferring  the plausible cosmological origin of a primordial GWB from its spectrum is notoriously difficult. 
For this reason, identifying robust features of the primordial spectrum is particularly important.
In this sense, the low-frequency spectrum of the signal (i.e.~$f<\fpk$) offers a unique opportunity, because its scaling is universal and does not depend on the details of the source \cite{Caprini:2009fx, Cai:2019cdl, Hook:2020phx}.
This range of the GW spectrum, on wavelengths longer than the correlation length of the source, is governed by causality and the propagation of GWs after their production.
We offer here a qualitative summary of these effects, and refer to \cite{Hook:2020phx} for a more detailed description.

For any GW source that has a finite time duration, its correlation length is necessarily finite (and smaller than the Hubble radius when the production ceases). 
All GW modes on wavelengths longer than this correlation length then are excited on short time scales (compared to their frequencies), with an amplitude that is the same for all frequencies. 
Intuitively, the source ``perceived'' by these low-frequency GW modes is an average of the microphysical source on many uncorrelated patches, and the $k$-dependence of the source on these long distances is fixed by causality to be the universal scaling of white noise.
As a result, the spectral shape of this so-called ``causality tail'' is $f^3$ for a RD universe, and can only be affected by variations of the EOS parameter or the presence of relativistic free-streaming particles, which modify the equation of motion of GWs by introducing an effective friction.
It is possible to obtain an approximate analytical expression for the causality tail as a function of the EOS parameter $w$ (under the assumption that it evolves slowly),
\begin{equation}
\OGW \propto f^\frac{1+15w}{1+3w} \quad (\text{slowly-varying }w)
\end{equation}
which reproduces $\OGW \propto f^3$ for RD, $\OGW \propto f$ for MD, and interpolates for the intermediate regimes.
The modification of the causality tail induced by $w\neq \tfrac 13$ are generically small, but for a loud GW background can be realistically detected \cite{Brzeminski:2022haa} even in the presence of astrophysical foregrounds \cite{Racco:2022bwj}, by exploiting the different spectral shapes of these backgrounds.

\begin{figure}[t]
\centering
\includegraphics[width=0.49\textwidth]{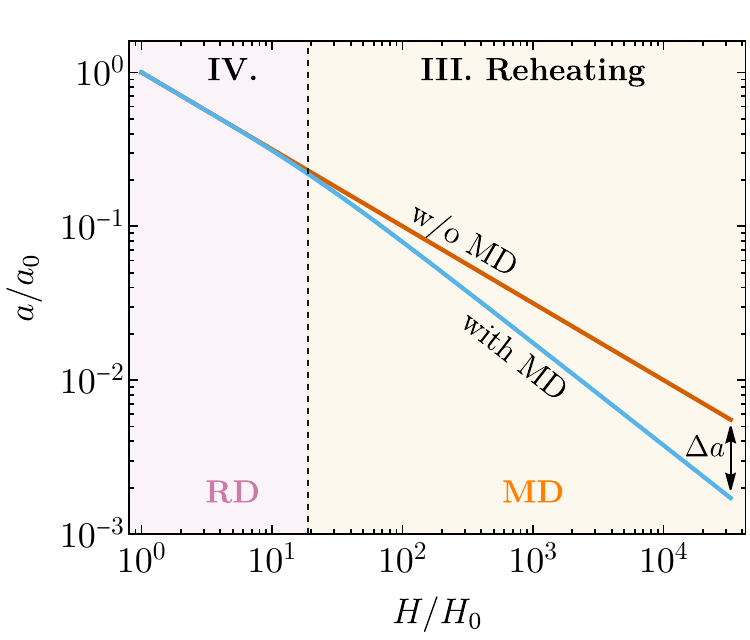}%
\hfill
\includegraphics[width=0.49\textwidth]{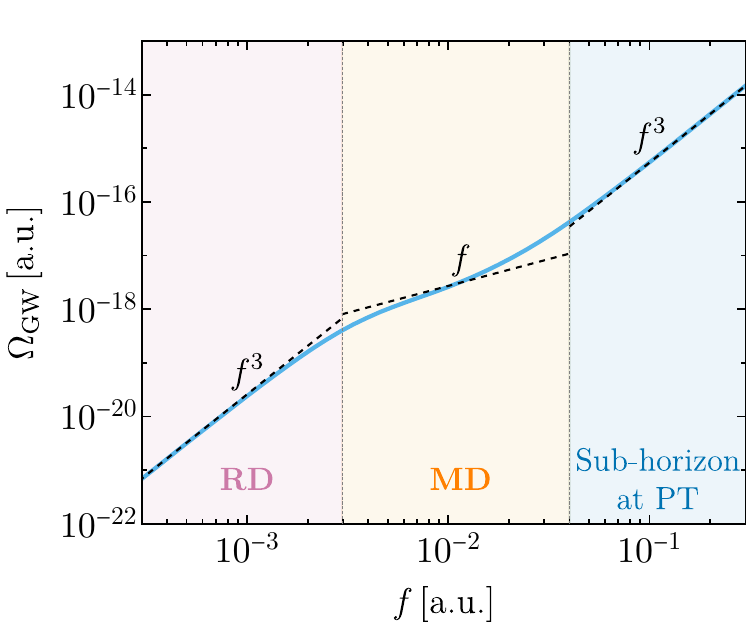} 
\caption{\textit{Left:} time evolution of the scale factor $a(t)$, compared for standard RD (orange) and with a MD reheating phase (blue). A reference Hubble value $H_0\equiv H(a_0)$ at late times is introduced. 
\textit{Right:} IR tail of the GW spectrum. The power law scaling of a given mode $f$ depends on the equation of state of the universe around horizon crossing. The evolution is determined by solving \cref{eq:reheating} with an inflaton decay rate $\gamma=0.01 \Hinf$.}
\label{fig:CT MD}  
\end{figure}
In this work, we solve numerically the equations of motion for long-wavelengths GWs as described in \cite{Hook:2020phx}, using the EOS obtained from a numerical evaluation of the system in \cref{eq:reheating}.
The resulting effect on the causality tail is shown in the right panel of \cref{fig:CT MD}. As expected, the power-law scaling observed for modes that re-enter the horizon during reheating gradually approaches a linear behaviour, eventually reconnecting with the cubic universal dependence expected for modes that are already sub-horizon at the time of production.

There is a further effect on the physical GW frequencies that is important to account for, when comparing the theoretical signal to the observational sensitivity.
If one fixes the temperatures of the universe today (as measured from the CMB) and at the start of the MD phase (which is the case if one assumes a particle model producing e.g.~SC), then extending the duration of the MD epoch implies a longer expansion history, as shown in the left panel of \cref{fig:CT MD}.
The values of $H$ on the left and right sides of the plot are fixed by the temperatures $T_0$, defined as an early time in RD from which the cosmological evolution follows $\LCDM$.
A longer MD phase corresponds to a faster expansion and a larger excursion of the scale factor by an amount $\Delta a$. 
Then, for any comoving momentum $k_\text{com}$, the physical frequency $k_\text{ph} = k_\text{com}/a$ is \textit{smaller} in the cosmological scenario with a longer MD.%
\footnote{We notice that the higher redshift of physical frequencies was not accounted for in e.g.~\cite{Gonstal:2025qky, Athron:2025pog}, who implicitly fix the ratio $a_0/a_\text{nuc}$ regardless of the duration of the MD phase, thus varying the exit temperature $\Tnuc$ from the SC.}
Graphically, the GW spectrum is shifted to the left (smaller frequencies), as shown in \cref{fig:GWB MD}.
The overall effect on the GWB of the MD phase is a suppression of the peak, and an amplification of the signal at low frequencies (as visible from the comparison of blue and orange curve).
The latter effect can ease the detection of the GWB in the low-frequency range.

We remark that the features of a MD phase (or generically $w<\tfrac 13$) reviewed in this section are an inevitable signature of the effects discussed in \cref{sec:DMsupercooled}.
A non-instantaneous reheating of the Universe after SC leads to both effects: a peculiar spectral shape of the causality tail of the GWB produced at the end of SC, and a modification of the relic abundance of DM candidates with respect to the generic dilution $(\Tnuc/\Tinf)^3$ if the DM abundance is set during II.
A future detection of a GWB with the features sketched in \cref{fig:GWB MD} would strongly suggest that the DM relic abundance was affected in the ways described in \cref{sec:DMsupercooled} for FO and FI.
Conversely, the detection of a DM candidate in a parameter space that looks un-natural for FI (as along the blue and red lines of \cref{fig:DDfreezein}, or below the typical values for a WIMP-like candidate as in \cref{fig:DirectSearch,fig:DirectSearchWIMP} would be a suggestive hint of a SC epoch. This hypothesis would corroborate the motivation to search for a primordial GWB generated at the end of SC.

Let us conclude by noting that the spectral shape of the CT depicted in \cref{fig:GWB MD}, due to an intermediate MD stage, would match more closely the preliminary indications about the spectral tilt $n_T\sim 1-2$
of the GWB detected at PTAs (as noted e.g.~in \cite{Franciolini:2023wjm}).
If future measurements of this GWB ended supporting the primordial hypothesis, a SC scenario with an extended reheating phase would be a motivated particle physics scenario.

\section{Conclusions}
\label{sec:conclusions}
In this paper, we consider the possibility that the Universe underwent a phase transition at early times with a stage of supercooling, where the latent heat dominates the energy budget before the completion of the PT.
The exponential dilution of pre-existing species, and the entropy damping at the end of SC, affect any dark sector and the dynamics of DM production at that time.
The final DM abundance is impacted by the generic dilution factor of \cref{eq:SCfactor}, and also by further effects that we clarify in our work if the freeze-out or (infra-red dominated) freeze-in is completed during the SC stage, and the reheating phase has a realistic duration, so that the Universe reheats to a temperature lower than at the start of SC (although the latter requirement is relaxed if the dark sector is very weakly coupled, or asymmetric).
The modified evolution of the Hubble rate (which stops its decrease at the onset of SC) anticipates the conclusion of FO and FI.
This implies smaller (for FO) and larger (for FI) couplings to achieve the present DM abundance, with respect to a cosmology without SC.

We illustrate for the first time how this affects the prospects for direct detection for DM.
For a WIMP-like scenario, SC brings natural candidates from excluded to allowed regions (\cref{fig:DirectSearchWIMP}). 
If the SC lasts more than a few $e$-folds (an interesting scenario realising this is an electroweak PT undergoing SC, and being terminated by the QCD PT), the effects on the parameter space for direct detection of DM produced by FO are more drastic (\cref{fig:DirectSearch}).
The opposite trend to FO is realised if DM is produced via FI, as usual: larger couplings are required, as we illustrate in \cref{fig:DDfreezein} for a millicharged DM candidate. The interesting implications are two-fold: the mass range above a TeV is closer to experimental reach; alternatively, the detection of DM on the standard FI line excludes a SC stage at temperatures below the DM mass.

Finally, we clarified that these effects, which arise if the reheating phase has a realistic finite duration, guarantee a further peculiar feature in the low-frequency (causality tail) of the gravitational wave background produced at the end of the PT. 
The stage of reheating features an EOS parameter closer to MD, so that the spectral tilt of the GWB at frequencies below the peak scales less steeply than $f^3$ (down to $f$ for pure MD), and the whole spectrum is shifted to lower frequencies because of the larger total redshift. 
These effects on the GWB and the DM parameter space are two predictions of the same dynamics of SC, and offer complementary handles on the physics of the early Universe. 

As future developments, it would be relevant to model more accurately the reheating phase, to get a more reliable prediction of the EOS parameter $w(a)$ and its implication for the GWB spectral shape. It would also be important to clarify if a prolonged stage of SC is indeed more tuned than previously believed \cite{Mishra:2024ehr, Agrawal:2025wvf}, as the effects that we describe become even more drastic for a long SC phase.
Another interesting direction (along the lines of \cite{Hambye:2018qjv}) would be to study the DM production during the reheating stage, which is relevant if the FO or FI is not completed at $\Trh$.

If our early Universe displayed a stage of supercooling, the prospects for direct detection of DM can be brighter than commonly thought.

\acknowledgments
We thank Admir Greljo and Toby Opferkuch for the valuable suggestion leading to \cref{sec:WIMP}.
A.S. would like to thank Toby Opferkuch also for other useful discussions.\\ 
D.R.~is supported at U.~of Zurich by the UZH Postdoc Grant 2023 Nr.\,FK-23-130.\\
A.S. is supported by the Italian Ministry of University and Research (MUR) and by the European Union’s NextGenerationEU program under the Young Researchers 2024 SoE Action, research project ‘SHYNE’, ID: SOE 20240000025.

\appendices

\section{Supercooling during the freeze-out of a cool dark sector}\label{app:darkQED}
There exist scenarios in which the portal between the dark and visible sectors is sufficiently weak that thermal production of DM, after reheating of the visible sector, is negligible or entirely absent. In such cases, the dark sector does not efficiently re-thermalize after the transition, and the relic abundance generated during supercooling can remain relevant. This opens the possibility for the dynamics illustrated in \cref{fig:SC_freezeout} to have a sizeable impact, even in the absence of a matter-dominated epoch and despite the consequent high reheating temperature $\Trh \approx \Tinf$.

Let a consider a hidden dark sector that is internally thermalized at a temperature $T_D$, related to the visible sector temperature by the parameter $\xi \equiv T/T_D$. The dark sector may interact with the Standard Model through a feeble portal interaction, but we assume it is insufficient to establish thermal equilibrium between the two sectors, so that $\xi$ remains approximately constant.  If this dark sector features a non-trivial spectrum with interacting species, some of them  may decouple from equilibrium and serve as viable dark matter candidates. In such a setup, one typically obtains a freeze-out scenario within the hidden sector, where the dark matter abundance is determined by the annihilations among dark species. However, the Hubble expansion rate that governs this freeze-out receives contributions from both the visible and hidden energy densities and, in the case of a supercooled phase transition, also from the vacuum energy component. This makes the dynamics sensitive to the thermal history of the visible sector, even if the dark sector is otherwise secluded. Explicitely, the Hubble rate would read,
\begin{equation}
H(T,T_D)^2= \frac{\frac{\pi^2}{30}(g_{*D} T_D^4 + g_* T^4)+\Delta V}{3\,m_{\text{Pl}}^2}=\frac{\frac{\pi^2}{30}(g_{*D}+\xi^4 g_*)T_D^4+\Delta V}{3\,m_{\text{Pl}}^2}\,.
\end{equation}

In a standard radiation-dominated universe, solving the Boltzmann equation within the hidden sector shows that, if $g_*^D \ll g_* \, \xi^4$ (which is the relevant case if the dark sector is colder than the SM), the final relic density can be estimated using the standard cold relic approximation with the rescaling $\Omega_{\text{DM}} h^2 \to \xi^{-1} \Omega_{\text{DM}} h^2$ \cite{Chu:2011be}.  If the thermal history also includes an epoch of supercooling, two main effects arise:
\begin{itemize}
    \item The supercooling phase during which the Hubble rate remains constant, from $\Tinf$ to $\Tnuc$ in the SM bath, spans in the dark sector a range of temperature from $\Tinf/\xi$ to $\Tnuc/\xi$, thereby shifting the supercooling scales that are relevant for the dark sector dynamics.
    
    \item The characteristic features of freeze-out during supercooling, as illustrated in \cref{fig:SC_freezeout}, are no longer confined to the temperature window $\Tinf$–$\Trh$, but extend down to $\Tnuc$, since reheating does not affect the hidden sector.
\end{itemize}

For concreteness, we mention a well-studied model of dark matter where this mechanism may be relevant. Consider a secluded dark sector consisting of a minimal dark Abelian $U(1)_D$ gauge symmetry with a single Dirac fermion. This dark sector can interact with the Standard Model via a kinetic mixing portal:
\begin{equation}
    \mathcal{L} \supset -\frac{\hat{\varepsilon}}{2} B^{\mu \nu} F_{\mu \nu}^{\prime}\,,
\end{equation}
where $B^{\mu\nu}$ is the SM hypercharge field strength tensor and $F_{\mu\nu}^{\prime}$ is the field strength of the dark photon. Key parameters of the model are the dark gauge coupling $   \alpha_{D} = e_D^{2}/4\pi$ and the mass of the dark fermion $ m_{\DM} $. As usual, one can rotate to a basis where the kinetic terms are canonically normalized. Following~\cite{Chu:2011be}, the relevant Lagrangian terms involving the SM photon $A$, the SM $Z$ boson, and the dark photon $A^{\prime}$ read:
\begin{equation}
\mathcal{L} \supset 
     -e J_{\text{EM}}^{\mu} A_{\mu}
    + e_{D} \frac{\hat{\varepsilon} \cos \theta_{\hat{\varepsilon}}}{\sqrt{1 - \hat{\varepsilon}^{2}}} J_{\text{DM}}^{\mu} A_{\mu}
    - e_{D} J_{\text{DM}}^{\mu} A_{\mu}^{\prime}
     - e_{D} \frac{\hat{\varepsilon} \sin \theta_{\hat{\varepsilon}}}{\sqrt{1 - \hat{\varepsilon}^{2}}} J_{\text{DM}}^{\mu} Z_{\mu}
    - g \frac{\cos \theta_W}{\cos \theta_{\hat{\varepsilon}}} J_{Z}^{\mu} Z_{\mu}\,,
\end{equation}
where $\theta_{\hat{\varepsilon}}$ is defined by $ \tan \theta_{\hat{\varepsilon}}=\tan \theta_{W} / \sqrt{1-\hat{\varepsilon}^{2}}$. With this choice, dark fermions are coupled to ordinary photon and $Z$ boson. For simplicity, we can further define $\varepsilon\equiv\hat{\varepsilon} \cos \theta_{\hat{\varepsilon}}/ \sqrt{1-\hat{\varepsilon}^{2}}$, so that relevant processes only depend on the combination $\kappa\equiv\varepsilon\sqrt{\alpha_D/\alpha} =\varepsilon \,e_D/e$, that we discussed in \cref{sec:supercool_freezein}.
Different ranges of the parameters $\kappa,\alpha_D,m_{\DM}$ lead to distinct regimes in which a viable DM relic can be obtained \cite{Chu:2011be}. 
While \cref{sec:supercool_freezein} focuses on the freeze-in scenario, we consider here the freeze-out regime achievable for larger values of $\alpha_D$.

In this case, the dark sector particles can interact fast enough to evolve in thermal equilibrium with a temperature $T_D$, and eventually dark fermions freeze out
through the annihilation channel $e^{\prime} e^{\prime} \rightarrow A^{\prime} A^{\prime}$, with a cross section proportional to $\alpha_{D}^{2} /m_{\DM}^2$. Crucially, after the completion of the supercooled transition in the visible sector, any subsequent thermal production of dark particles, even if the SM temperature is $\Trh > T_{\text{f.o.}}$, will be strongly suppressed due to the smallness of the mixing parameter $\varepsilon$.

\section{Baryogenesis and Asymmetric Dark Matter}
\label{app:baryogenesis}

Supercooling leads to a significant dilution of particle number densities, provided they are not thermally regenerated during reheating. In particular, asymmetries between particle and antiparticle number densities are neither enhanced nor erased if the scalar degrees of freedom only reheats a symmetric component. This makes supercooling a compelling mechanism to account for the smallness of the baryon asymmetry of the Universe (BAU), as previously noted, e.g., in \cite{Baratella:2018pxi}:
\begin{equation}\label{eq:baryon_today}
    \eta_{\B}=\frac{n_{\B}-n_{\bar{\B}}}{s}=\eta_{\B}^{(0)} \cdot  \left(\frac{\Tnuc}{T_\textrm{inf}}\right)^3\cdot \left(\frac{T_{\text{RH}}}{T_{\text{inf}}}\right)=(8.7\pm0.1)\cdot 10^{-11}\,,
\end{equation}
where $\eta_B^{(0)}$ denotes the initial baryon asymmetry generated in the early Universe. Compared to standard baryogenesis scenarios, this cosmological history allows for a much larger initial asymmetry and potentially broadens the range of viable baryogenesis mechanisms\footnote{Similar ideas were suggested in the context of primordial inflation \cite{Krnjaic:2016ycc}. However, large initial asymmetries would lead to baryon isocurvature perturbations that are already excluded by cosmic microwave background observations \cite{Murai:2023ntj}.}. 
Remarkably, in the specific case of a supercooled electroweak transition terminated by the QCD transition, the asymmetry is parametrically set by
\begin{equation}
    \eta_b\simeq \eta_b^{(0)}\cdot\left(\# \frac{v_\textsc{ew}}{\LQCD}\right)^3\,,
\end{equation}
so that the smallness of the baryon asymmetry is directly related to the ratio of fundamental scales arising within the SM. 
On the other hand, the large initial asymmetry $\eta_b^{(0)}$ constrains the baryogenesis mechanism operating at higher scales. 

\begin{figure}[t!]
\centering
\includegraphics[width=0.75\textwidth]{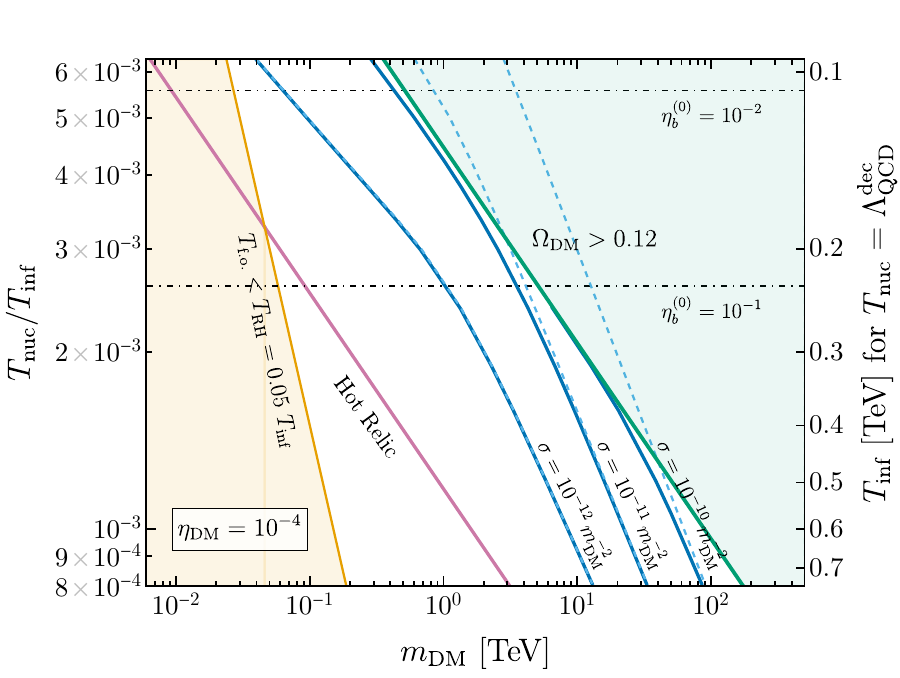}
\caption{The solid blue lines show the amount of supercooling required to achieve the correct DM abundance (as a function of $m_\DM$) for various annihilation cross sections $\sigma$, when an initial DM asymmetry $\eta_{\DM}=10^{-4}$ is present. The green line stresses that the freeze-out yield cannot be smaller than approximately $Y_{\text{f.o.}} \sim \eta_{\DM}$. The dashed lines represent the results that we would obtain if the asymmetry $\eta_{\DM}$ was vanishing (i.e.~same as solid lines in \cref{fig:DMrelic}). Horizontal dot-dashed lines illustrate the required initial baryon asymmetry $\eta_b^{(0)}$ assuming that there is no baryogenesis after the supercooling.}
\label{fig:DMrelicASYM}  
\end{figure}

It is also conceivable that the baryon asymmetry is generated after the supercooled transition. In this case, one avoids the need for large initial asymmetries and the risk of washout due to entropy injection. However, producing an asymmetry at low energies could also be challenging \cite{Bagherian:2025puf}. Finally, we note that the FOPT dynamics itself, especially with runaway bubbles, offer viable mechanisms for the production of a baryon asymmetry \cite{Konstandin:2011ds,Katz:2016adq,Cataldi:2024pgt,Azatov:2021irb,Dichtl:2023xqd}.

A compelling alternative explanation for the observed DM abudance in the universe involves the existence of an asymmetry between dark particles and their antiparticles \cite{Petraki:2013wwa,
Zurek:2013wia}, analogous to the baryon asymmetry responsible for the visible matter abundance.

The presence of a DM asymmetry can alter the conclusions drawn in \cref{sec:DMsupercooled}. In particular, the introduction of an asymmetry:
\begin{equation}
    \eta_{\DM}\equiv Y_{\DM} - Y_{\overline{\DM}} \,,
\end{equation}
implies that the freeze-out yield cannot be smaller than approximately $Y_{\text{f.o.}}\sim \eta_{\DM}$. This occurs because particle-antiparticle annihilation becomes inefficient, leaving the relic abundance primarily determined by the net particle excess rather than by the freeze-out of a symmetric population. 
We illustrate this behaviour in \cref{fig:DMrelicASYM} for the representative choice $\eta_{\DM}=10^{-4}$, assumed here to be independent of the supercooling dynamics and unrelated to the baryon asymmetry. We solve the coupled Boltzmann eq.s for DM particles and antiparticles as in \cite{Graesser:2011wi}. The green curve indicates the boundary in parameter space where we expect that the asymmetric component becomes dominant in determining the final DM abundance. To the right of this limit, DM is overproduced due to the large asymmetric contribution.

An interesting aspect of asymmetric DM models is the potential for a common origin with the baryon asymmetry, and therefore an explanation to the baryon-DM coincidence $\Omega_{\DM}\sim 5 \,\Omega_b$. If we assume a direct relation of the form $\eta_{\DM}=\kappa \,\eta_b^{(0)}$, and that $Y_{\DM}\simeq \eta_{\DM}\gg Y_{\overline{\DM}}$, we recover the standard estimate for the DM particle mass:
\begin{equation}
    m_{\DM}\simeq 5 \kappa^{-1} m_p \simeq 5 \kappa^{-1} \GeV\,.
\end{equation}
This relation arises from the fact that the entropy produced at the end of the supercooling era dilutes both dark matter and baryons equally, preserving their ratio. 

After the phase transition concludes, a reheating stage follows. If DM production during reheating is purely symmetric (equal numbers of DM particles and antiparticles), the previously diluted asymmetry remains unchanged, and the estimates derived above continue to provide reliable predictions, given that there exists an efficient annihilation channel for the symmetric component. On the other hand, if DM particles are long-lived or if the asymmetric component is further enhanced during reheating, the simple
picture outlined above must be revised.

\bibliographystyle{JHEP}
\bibliography{biblio}


\end{document}